# Ultra-precise determination of thicknesses and refractive indices of optically thick dispersive materials by dual-comb spectroscopy

Kana A. Sumihara[1], Sho Okubo[2], Makoto Okano[1], Hajime Inaba[2] and Shinichi Watanabe[1,*]

[1] *Department of Physics, Faculty of Science and Technology, Keio University, 3-14-1 Hiyoshi, Kohoku-ku, Yokohama, Kanagawa 223-8522, Japan.*

[2] *National Metrology Institute of Japan (NMIJ), National Institute of Advanced Industrial Science and Technology (AIST), Tsukuba, Ibaraki 305-8563, Japan.*

*Corresponding author (e-mail: watanabe@phys.keio.ac.jp)

**Precise measurements of the geometrical thickness of a sample and its refractive index are important for materials science, engineering, and medical diagnosis. Among the possible non-contact evaluation methods, optical interferometric techniques possess the potential of providing superior resolution. However, in the optical frequency region, the ambiguity in the absolute phase-shift makes it difficult to measure these parameters of optically thick dispersive materials with sufficient resolution. Here, we demonstrate that dual frequency-comb spectroscopy can be used to precisely determine the absolute sample-induced phase-shift by analyzing the data smoothness. This method enables simultaneous determination of the geometrical thickness and the refractive index of a planar sample with a precision of five and a half digits and an ultra-wide dynamic range. The thickness and the refractive index at 193.414 THz of a silicon wafer determined by this method are 0.52047(3) mm and 3.4756(3), respectively, without any prior knowledge of the refractive index.**

The geometrical thicknesses $d$ and the complex refractive indices of layers ($n(\omega) + ik(\omega)$, where $\omega$ is the optical angular frequency) are fundamental parameters for materials science, and thus ultra-precise determination of these parameters in a wide range of materials is important. Due to advances in optical technologies, several techniques can be used to implement precision metrology[1,2,3,4,5,6]. For example, laser interferometry has been used for the dimensional metrology of optically thin samples (whose thickness is comparable with the wavelength of light, $\lambda$)[7]. In this method, $d$ and $n(\omega)$ of a planar sample can be determined by analyzing the absolute sample-induced phase change $\Delta\Phi_{abs}(\omega)$. However, in conventional interferometry of optically thick samples (i.e. $d > \lambda$), we need to consider the ambiguity in the value of $\Delta\Phi_{abs}(\omega)$[8], because in this case, $\Delta\Phi_{abs}(\omega)$ is usually not identical to the measured phase change $\Delta\phi_{exp}(\omega)$ but can differ by multiples of $2\pi$:

$$\Delta\Phi_{abs}(\omega) = 2\pi M(\omega) + \Delta\phi_{exp}(\omega) \qquad (1)$$

where $M(\omega)$ is an (unknown) integer. Because $M(\omega)$ can be used to evaluate $d$ and $n(\omega)$ of a planar sample, various interferometric methods have been proposed to precisely determine $M(\omega)$[8], for instance interferometric methods that employ rotating[9,10,11] or moving[12,13] the sample. However, since the mechanical system that is required for such sample translations, causes additional measurement uncertainties, an ultra-precise determination of $d$ and $n(\omega)$ by these methods is still difficult. Moreover, optical ellipsometry is usually only applicable in the case of optically thin samples[14,15,16,17], and low-coherence interferometry can only be used to measure the group refractive index ($n_g(\omega)$) of the sample[18,19,20,21]. Thus, a more sophisticated optical technique is required.

The development of the optical frequency comb (OFC) technology has revolutionized the field of precision metrology. Dual-comb spectroscopy (DCS), which uses two OFCs[22,23,24,25,26,27], enables us to unambiguously determine $\Delta\Phi_{abs}(\omega)$ in open-air ranging applications[28,29,30,31] by combining time-of-flight and interferometric measurements, resulting in an ultra-wide dynamic range of the determination of the absolute distance. This concept has been applied to materials science in order to determine $d$ and $n_g(\omega)$ of samples like wafers[19,20,32,33]. However, because the distortion of the optical wave-packet obstructs the determination of $\Delta\Phi_{abs}(\omega)$ due to the dispersion of $n(\omega)$, the same precision as in open-air ranging has not been achieved.

Here, we demonstrate the ultra-precise determination of $d$, $n(\omega)$ of an optically thick dispersive material with an ultra-wide dynamic range. We determine $\Delta\Phi_{abs}$ by analyzing the smoothness of the $n(\omega)$ and $k(\omega)$ spectra derived from the experimental data (i.e. the spectra should exhibit no artificially induced oscillating behavior with respect to frequency). We show that $d$ and $n(\omega)$ of a silicon wafer can be simultaneously determined with a precision of five and a half digits: $d=520473.7\pm1.5$ nm and $n=3.47563\pm0.00001$ at 193.414 THz with a relative uncertainty of $\sim3\times10^{-6}$. This precision of $n(\omega)$ is the best value that has so far been reported for a planar sample, and is comparable with the precision of the interferometric measurement of prism-shaped samples[12]. Moreover, the value of $n(\omega)$ is consistent with the literature value within the previously reported uncertainty.

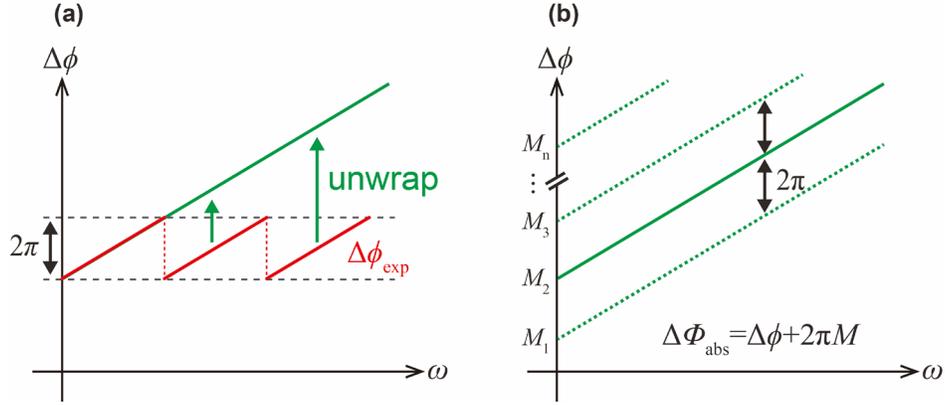

**Figure 1 | Phase ambiguity in DCS**. **a,** Schematic of $\Delta\phi_{exp}(\omega)$ obtained in a DCS experiment (red lines). Because the range of $\Delta\phi_{exp}$ is restricted to 0–2π, $\Delta\phi_{exp}(\omega)$ is discontinuous and the recorded data is divided into locally continuous data segments. Due to the ultra-high frequency resolution of DCS, a continuous $\Delta\phi(\omega)$ curve can be obtained by unwrapping the data segments starting from a selected data segment (green line). **b,** As long as the value of $M(\omega)$ of the starting segment of the unwrapping procedure, which is defined as $M$, is not known exactly, there are several possible candidates for $\Delta\Phi_{abs}$ (green lines distinguished by the different possible values of $M$ denoted by $M_1, M_2, \ldots, M_n$). To determine $\Delta\Phi_{abs}(\omega)$, we have to determine $M$.

**Results**

**The criterion for the correct value of *M*.**

Figure 1 describes the underlying problem in the DCS measurement; the phase ambiguity of the experimental $\Delta\phi_{exp}(\omega)$ data (Fig. 1a) and the resulting possible candidates for $\Delta\Phi_{abs}(\omega)$ (Fig. 1b). Here, we explain how to solve this problem of ambiguity in $\Delta\Phi_{abs}(\omega)$ obtained by DCS. As a general approach, we first construct a continuous $\Delta\phi(\omega)$ curve (Fig. 1a) by unwrapping the data segments starting from a selected data segment, and define the $M(\omega)$ of this starting segment as $M$. Therefore, we consider the equation $\Delta\Phi_{abs}(\omega)=\Delta\phi(\omega)+2\pi M$ with $M$ being unknown. Our solution is based on a single simple criterion that states that the complex refractive index spectrum $N(\omega)=n(\omega)+ik(\omega)$ is the smoothest when we derive them from the experimental data using the correct values of $M$ and $d$. When the $n(\omega)$ and $k(\omega)$ spectra are calculated with incorrect values of $M$ and $d$, these spectra contain an artificial oscillating component. This oscillating behavior with respect to frequency is caused by the incorrect analysis of the effect of the multiple internal reflections in $\Delta\phi_{exp}(\omega)$ and in the power transmittance $T(\omega)$. Since the complex refractive indices of most naturally occurring materials do not exhibit any oscillating behavior with respect to frequency, the smoothness of $n(\omega)$ and $k(\omega)$ can be used as an indicator of the correctness of $M$ and $d$.

Figure 2 shows the flowchart of our analytical procedure. First, we measure the interferograms with and without the sample, $U_{w/}(t)$ and $U_{w/o}(t)$, respectively (Fig. 2a). The details of the interferogram measurement are explained in the Methods section. Then, by Fourier transform (FT) of each interferogram, $\Delta\phi(\omega)$ and $T(\omega)$ are derived (Fig. 2b). Here, $\Delta\phi(\omega)=\phi_{w/}(\omega)-\phi_{w/o}(\omega)$ and $T(\omega)=A_{w/}(\omega)/A_{w/o}(\omega)$, where $\phi_{w/}(\omega)(\phi_{w/o}(\omega))$ and $A_{w/}(\omega)(A_{w/o}(\omega))$ are the phase and the power amplitude of $U_{w/}(t)(U_{w/o}(t))$, respectively. $N(\omega)$ can be calculated from $T(\omega)$, $\Delta\phi(\omega)$, $M$, and $d$ by solving the analytical equations (10) and (11) in the Methods section for numerous pairs of $(M_j,d_j)$. The pairs $(M_j,d_j)$ describe the considered candidates for the correct combination of $M$ and $d$, where $j$ runs over all elements of a two-dimensional array that consists of all possible combinations of the candidate values of $M$ and $d$. To evaluate the smoothness of $n(\omega)$ and $k(\omega)$ derived for each pair $(M_j,d_j)$, we use the following two functions $E_n(M,d)$ and $E_k(M,d)$ (See Methods for details):

$$E_n(M,d) = \sum_i \left| \left( \left.\frac{d^2 n}{d\omega^2}\right|_{\omega_{i+1}} - \left.\frac{d^2 n}{d\omega^2}\right|_{\omega_i} \right) \right| \qquad (2)$$

$$E_k(M,d) = \sum_i \left| \left( \left.\frac{d^2 k}{d\omega^2}\right|_{\omega_{i+1}} - \left.\frac{d^2 k}{d\omega^2}\right|_{\omega_i} \right) \right| \qquad (3)$$

Here, the subscript $i$ identifies the discrete frequency values of the FT data in the range from the lowest to the highest frequency in the observable spectral region. The correct $M$ and $d$ values are those that minimize $E_n(M,d)+E_k(M,d)$ (Fig. 2d).

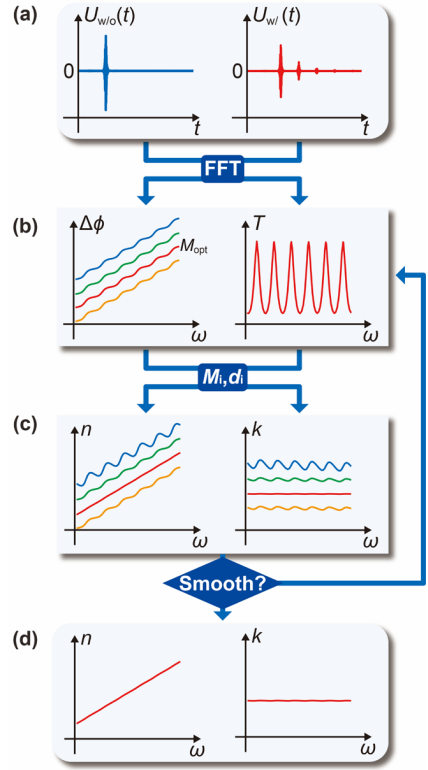

**Figure 2 | Flow chart of the analytical procedure. a,** The interferograms with and without the sample, $U_{w/}(t)$ and $U_{w/o}(t)$. $U_{w/}(t)$ contains the echo signals due to the multiple internal reflections. **b,** $\Delta\phi(\omega)$ and $T(\omega)$ calculated by Fourier transform of each interferogram. $\Delta\phi(\omega)$ and $T(\omega)$ show oscillating behaviors with respect to frequency due to multiple internal reflections. The colored curves visualize the ambiguity in the value $\Delta\Phi_{abs}(\omega)$. **c,** By using $T(\omega)$, $\Delta\phi(\omega)$, $M_j$, and $d_j$, $n(\omega)$ and $k(\omega)$ are calculated for each pair $(M_j,d_j)$ as shown by the curves with different colors. **d,** We consider the pair $(M_j,d_j)$ that minimize the value of $E_n(M,d)+E_k(M,d)$ (explained in the text) as the correct values.

Note that a similar analytical procedure has been proposed to determine $d$ in the terahertz frequency range[34,35], and this idea has been applied to the characterization of various materials[36,37,38,39]. However, in the terahertz frequency range, the value of $M$ is uniquely and easily determined by extrapolating $\Delta\phi(\omega)$ to $\omega=0$, where $\Delta\Phi_{abs}$ should be always zero. This distinct difference between the determination of $M$ and $d$ at terahertz and at optical frequencies makes the determination of $N(\omega)$ in the optical frequency range very difficult.

**Measurement system.**

Figure 3 shows our DCS system consisting of two OFCs (S-comb and L-comb) (See Methods for details). In this paper, we used an undoped silicon wafer with a nominal thickness of 525±25 μm as sample. The complex refractive index of silicon is frequency dependent around 192.5–197.5 THz[40].

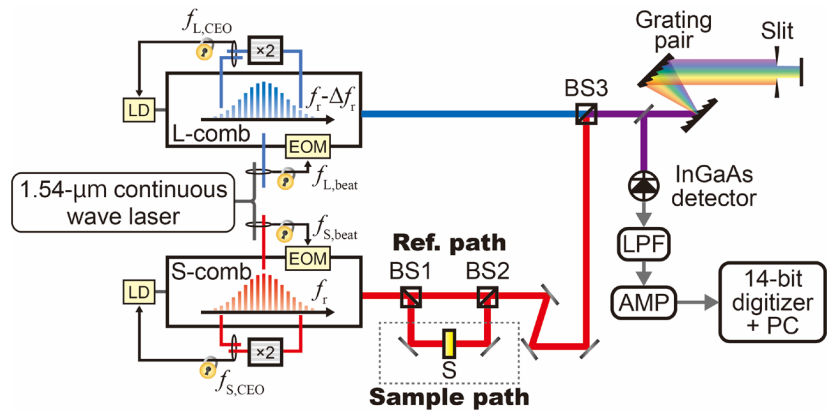

**Figure 3 | Experimental DCS setup.** The S-comb beam (red line) is divided into two paths by the beam splitter BS1. The paths for the transmitted and reflected beams are referred to as the reference (ref.) and sample paths, respectively. The beam that enters the sample path, is reflected by a mirror, passes through the sample, is reflected by another mirror, and is again combined with the beam from the ref. path by BS2. The combined S-comb beam is combined with the L-comb beam (blue line) by BS3. After that, the bandwidth of the combined beam (192.5–197.5 THz) is spatially filtered by a grating pair and an optical slit to avoid aliasing effects. Finally, the beam is detected by an InGaAs detector. The signal is electrically filtered by a low-pass filter (LPF), amplified by an amplifier (AMP), and is sampled by a 14-bit digitizer. LD: laser diode, EOM: electro-optic modulator, PC: personal computer.

We mounted the sample on a mirror mount and fixed this holder on a motorized translation stage. By moving the translation stage, we can easily obtain the interferograms with and without the sample. The direction of the optical beam was carefully aligned to achieve normal incidence on the sample surface within an uncertainty of ≈5 mrad. All measurements were performed in ambient atmosphere at room temperature (23±2 °C).

**Analysis of the interferogram data.**

Figure 4a shows a typical averaged interferogram obtained with the sample, $U_{w/}(t)$ (time used for averaging: ≈ 0.87 s, see Methods). The peak around 10.3 (11.2) ns corresponds to the interference signal between the L-comb beam and the S-comb beam that has passed through the ref. (sample) path. Figure 4b shows an enlarged view of the interference signals related to the sample path for two conditions: with (red data, $U_{w/}$) and without the sample (blue data, $U_{w/o}$). The interference signal in $U_{w/}(t)$ has several equally-spaced echo signals, which are a result of the multiple reflections inside the sample[32]. The averaged $T(\omega)$ and the $\Delta\phi(\omega)$ spectra derived from the interferogram data are plotted in Fig. 4c and 4d, respectively. We start the unwrapping procedure for $\Delta\phi(\omega)$ at 193.414 THz where $\Delta\phi_{exp}(\omega)$ is close to 0 rad. Note that, in this work, we calculated $T(\omega)$ and $\Delta\phi(\omega)$ using the interference information related to both the sample path and the ref. path (see Methods). $T(\omega)$ and $\Delta\phi(\omega)$ were averaged 1833 times, where each $T(\omega)$ and $\Delta\phi(\omega)$ spectrum for the averaging procedure was determined from one dataset consisting of $U_{i,w/}(t)$ and $U_{i,w/o}(t)$ (the definition of a dataset is provided in the Methods). The important feature of the $T(\omega)$ and $\Delta\phi(\omega)$ spectra is that both show an oscillating behavior with respect to frequency, corresponding to the etalon fringes from multiple internal reflections.

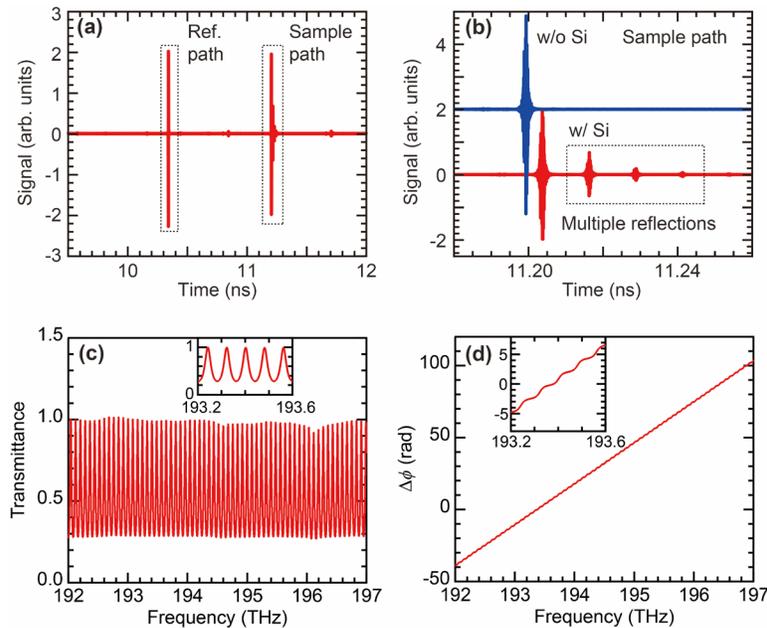

**Figure 4 | Interferogram data and the derived spectra of transmittance and the phase difference. a,** Typical averaged interferogram including the information of the sample, $U_{w/}(t)$. **b,** Typical averaged interferogram data in the vicinity of the interference signals related to the sample path in the cases with (red) and without (blue) the sample. Data are offset for clarity. **c,** Averaged power transmittance $T(\omega)$ and **d,** unwrapped phase difference, $\Delta\phi(\omega)$, derived from $U_{w/}(t)$ and $U_{w/o}(t)$.

Hereafter, we explain the protocol used to precisely determine $N(\omega)$. In this paper, we prepared about 25 million pairs of $(M_j,d_j)$ ($251 \leq M_j \leq 1500$; 515 μm $\leq d_j \leq$ 535 μm with a discretization of 0.001 μm). We calculated $n(\omega)$ and $k(\omega)$ for each pair $(M_j,d_j)$ and evaluated $E(M,d)=E_n(M,d)+E_k(M,d)$. Figure 5a shows the two-dimensional colormap of $E(M,d)$ on a logarithmic scale. $E(M,d)$ has a clear global minimum at $M=831$ and $d=520.474$ μm. Figure 5b plots the line profile of $E(M,d)$ at $M=831$ as a function of $d$, and the oscillation of $E$ observed in this figure has a period of $\lambda_0/2n_{air}$ (see Supplementary Information), where $\lambda_0$ and $n_{air}$ are the center wavelength of the interference signal and the refractive index of air in the measurable range, respectively. $E(M,d)$ at $M=831$ exhibits a global minimum at $d=520.474$ μm, and the values of the local minima gradually increase as the absolute difference between $d$ and $520.474$ μm becomes larger. Figure 5c plots the line profile of $E(M,d)$ at $d=520.474$ μm as a function of $M$; this curve has a minimum at $M=831$ and increases monotonically as the absolute difference between $M$ and 831 becomes larger. Figure 5d shows $E(M,d)$ in the vicinity of the global minimum.

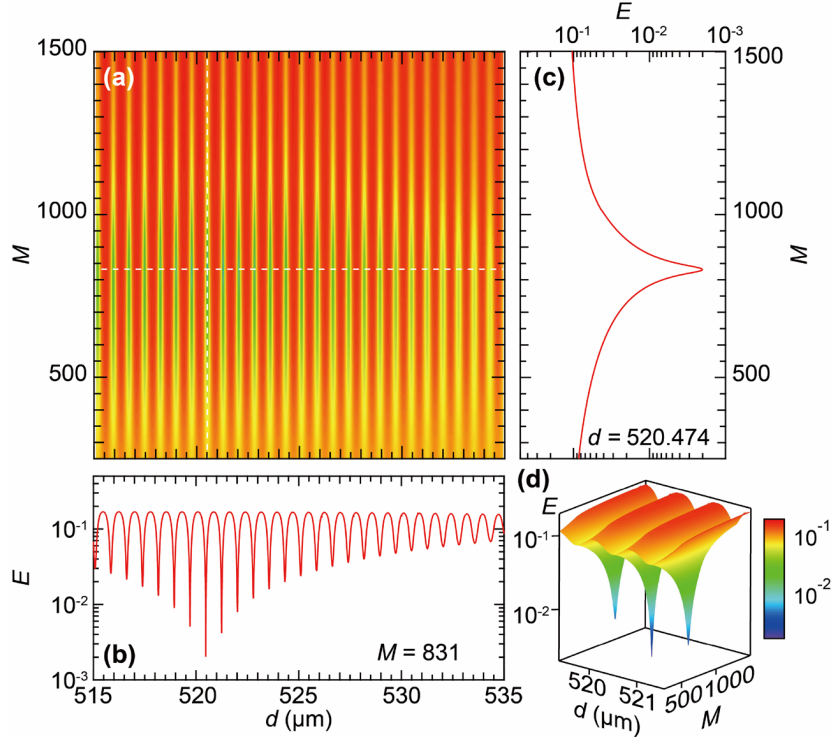

**Figure 5 | Demonstration of determination of *M* and *d*. a,** The two-dimensional (2D) colormap of $E(M,d)$, which is used to evaluate the smoothness of $n(\omega)$ and $k(\omega)$, on a logarithmic scale. **b,** Cross section of the data in (a) at $M=831$. **c,** Cross section of the data in (a) at $d=520.474$ μm. **d,** 3D image of $E(M,d)$ in the vicinity of the global minimum.

To evaluate the precision of our evaluation method, we estimated the values of the standard deviation of $M$ and $d$ for various averaging conditions of $T(\omega)$ and $\Delta\phi(\omega)$. We divided our 5500 spectra of $T(\omega)$ and $\Delta\phi(\omega)$ into subsets consisting of either $N_{\text{ave}}=$ 125, 250, 500, 1100, 1375, or 1833 spectra. Then, we averaged $T(\omega)$ and $\Delta\phi(\omega)$ over each subset for a selected $N_{ave}$ to determine the mean and the standard deviation of $M$ and $d$ (for example, $N_{ave}$=1375 allows us to determine the standard deviation using four data points for $M$). Figure 6a and 6b show the $N_{ave}$ dependences of $M$ and $d$, respectively. As shown in Fig. 6a, the averaged $M$ values are always closest to 831, independent of $N_{ave}$. In addition, the uncertainty in $M$ is smaller than ±1 for $N_{ave}$=1100 and the uncertainty in $M$ vanishes for $N_{ave}$=1375 because four determined $M$ have the same value of 831. This means that we can determine $M$ uniquely when $N_{ave} \geq 1375$, and thus we can precisely determine $\Delta\Phi_{abs}(\omega)$. In Fig. 6b, the mean value of $d$ is 520.473–520.474 µm and its error bar is less than 5 nm for all considered $N_{ave}$ values. The error bar gradually becomes smaller for larger $N_{ave}$ values, and it becomes less than 1.5 nm for $N_{ave} \geq 1100$. Our obtained value is consistent with the nominal value of the sample (525±25 μm), supporting the validity of our method.

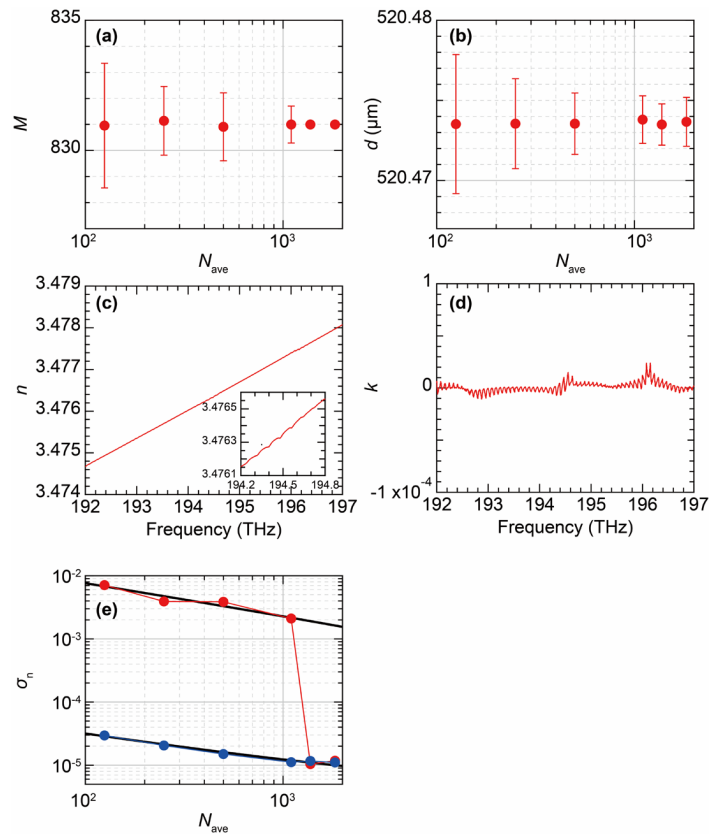

**Figure 6 | Discussion of the precision.** The $N_{ave}$ dependence of (a) $M$ and (b) that of $d$. The error bars represent the standard deviation. (c) $n(\omega)$ and (d) $k(\omega)$ of the undoped silicon wafer for $N_{ave}$=1833. (e) $N_{ave}$ dependence of the standard deviation of $n$ at 193.414 THz including the uncertainty of $M$ (red dots) and calculated using $M$=831 without uncertainty (blue dots). The black lines are proportional to $N_{ave}^{-1/2}$.

**The complex refractive index of silicon**

Since $d$ in our experiment was determined precisely when $N_{ave}=1833$, we consider the $n(\omega)$ and $k(\omega)$ calculated from the three pairs of $T(\omega)$ and $\Delta\phi(\omega)$ obtained for $N_{ave}=1833$. The averaged $n(\omega)$ and $k(\omega)$ are plotted in Fig. 6c and 6d, respectively. It is found that both $n(\omega)$ and $k(\omega)$ are rather smooth despite the oscillating behavior of the underlying data shown in Fig. 4c and 4d. This smoothness indicates that we correctly accounted for the effect of multiple internal reflections. In addition, $n(\omega)$ increases with frequency (normal dispersion), while $k(\omega)$ is almost zero. Because these behaviors are consistent with the literature[40], it is proven that our evaluation method is valid.

As shown in the inset of Fig. 6c and in Fig. 6d, even though the amplitudes of the oscillating component in the $n(\omega)$ and $k(\omega)$ spectra are minimized, a small residual oscillating component with an amplitude on the order of $\sim 10^{-5}$ is still observed. The origin of this small component is considered to be the uncertainty of $T(\omega)$. In contrast to $k(\omega)$, the uncertainty of $T(\omega)$ has only a weak influence on the uncertainty of $n(\omega)$, since $n(\omega)$ does not directly depend on $T(\omega)$ as shown in equation (10).

The red dots in Fig. 6e show the values of the standard deviation of $n(\omega)$ at 193.414 THz, $\sigma_n$, obtained for different $N_{ave}$ values using the corresponding standard deviation of $M$, $\sigma_M$. $\sigma_n$ gradually decreases in the range of $N_{ave}=125$–1100 and it dramatically decreases between $N_{ave}=1100$ and 1375 (more than 100 times smaller) because the value of $M$ is uniquely determined for $N_{ave} \geq 1375$. The blue dots in Fig. 6e correspond to the value of $\sigma_n$ calculated using $M=831$ and $\sigma_M=0$. In this case, $\sigma_n$ gradually decreases in the entire range of $N_{ave}$ and seems to approach a constant value for $N_{ave} \geq 1375$. More details of $\sigma_n$ are provided in the Discussion.

From Figs. 6c and 6e, we find that $n=3.47563\pm0.00001$ at 193.414 THz for $N_{ave}=1833$. This value is consistent with the refractive index of silicon at 193.414 THz reported previously in ref. [40], $n=3.4757\pm0.002$. This quantitative agreement proves that our method can be used to determine $n(\omega)$ precisely.

## Discussion

Firstly, we discuss the impact of the unique determination of $M$ on the precision of $n(\omega)$. Because the leading term of $n$ in equation (10) is $2\pi M$, it is considered that $\sigma_n$ is mainly determined by $\sigma_M$ as long as $M$ has not been determined uniquely; $\sigma_n \sim (2\pi\sigma_M c)/d\omega$. For the presently used experimental conditions, $\sigma_n$ is evaluated to be $3\times10^{-3}\sigma_M$, which governs the degree of uncertainty of the red dots in Fig. 6e for $N_{ave} \leq 1100$. Once $M$ has been uniquely determined, $\sigma_n$ dramatically decreases by more than two orders of magnitude. This drastic decrease in $\sigma_n$ is the main impact of the unique determination of $M$.

Secondly, we discuss the $N_{ave}$ dependence of $\sigma_n$ using the uniquely determined $M$ (Fig. 6e; blue dots). When $\sigma_M=0$, $\sigma_n$ is proportional to the uncertainty of $\Delta\phi_{exp}(\omega)$, $\sigma_{\Delta\phi(\omega)}$ (See Supplementary Information). In our experiment, $\sigma_{\Delta\phi(\omega)}$ is proportional to $N_{ave}^{-0.5}$ for $N_{ave} \leq 1100$ and tends to be constant for $N_{ave}>1100$ ($\sigma_{\Delta\phi(\omega)}\sim10$ mrad, data not shown). This lower limit of $\sigma_{\Delta\phi(\omega)}$ determines the lower limit of $\sigma_n$, which is $1.2\times10^{-5}$. Furthermore, the standard deviation of $d$, $\sigma_d$, does also not decrease further when $N_{ave}>1100$ (see Fig. 6b). This indicates that the lower limits of $\sigma_n$ and $\sigma_d$ in our experiment are not determined by the stability of our DCS system, but other experimental uncertainties due to changes in environmental conditions (such as temperature, pressure, and humidity), which change $n_{air}$ and $n(\omega)$. For instance, a temperature variation of 0.3 K causes a change in the refractive index of silicon on the order of $5.6\times10^{-5}$, which may limit $\sigma_n$. A discussion of the contributions to the experimental uncertainty is provided in the Supplementary Information.

Thirdly, we discuss why we were able to precisely determine $d$ even at $N_{ave}\leq1100$ where $\sigma_M\neq 0$. As mentioned above, the local minima of $E(M,d)$ appear at an interval of $\lambda_0/2n_{air}$ when plotted as a function of $d$. As the slope of $E(M,d)$ is quite steep around each local minimum (see Fig. 5b), $\sigma_d$ becomes much smaller than $\lambda_0/2n_{air}$ (~0.7 μm) even when we choose one of these local minima instead of the global minimum. In addition, because $\lambda_0/2n_{air}$ is larger than the uncertainty of $d$ for $N_{ave}=125$ in our experiment, we can choose the correct minimum, i.e., the global minimum, for all values of $N_{ave}$ as shown in Fig. 6b.

Finally, we discuss the experimental requirements for a unique determination of $M$ by our method. Our analytical protocol considers the smoothness of $N(\omega)$ to decide whether the values of $d$ and $M$ are correct. An $N(\omega)$ spectrum derived for an certain pair $(M_j,d_j)$ can contain an artificial oscillating component due to the actual interference between multiple-reflected optical pulses inside the sample. This means that the period of the artificial oscillating component in the derived $N(\omega)$ spectrum is determined by the optical-path difference between two adjacent optical pulses where one of these pulses is the result of multiple internal reflection. This optical-path difference is equal to $2n(\omega)d$. Because $\sigma_d$ is small even at smaller $N_{ave}$ values as mentioned above, the uncertainty of $2n(\omega)d$ is governed by $\sigma_n$ when $\sigma_M\neq 0$. When the value of $M$ changes by +1, the value of $n(\omega)$ in our experiment changes on the order of 0.003, which causes a change in the period of the artificial oscillating component in $n(\omega)$ by 0.09%. As a result, the period of the oscillating behavior of $\Delta\phi(\omega)$ with respect to frequency that is inversely estimated from the calculated $n(\omega)$, is not consistent with the measured oscillation period of $\Delta\phi(\omega)$. Because of this difference, the difference between the phases of the inversely calculated $\Delta\phi(\omega)$ and the measured $\Delta\phi(\omega)$ increases by ~0.6 mrad per oscillation. Because the number of oscillations in the measured $\Delta\phi(\omega)$ spectrum is ~60 within the measured frequency region (see Fig. 4d), the maximum phase difference becomes ~36 mrad. Thus, when $\sigma_{\Delta\phi(\omega)}$ is smaller than ~36 mrad, it is possible to uniquely determine $M$. By accumulating the data at each frequency, we verified that $\sigma_{\Delta\phi(\omega)} \approx 10$ mrad. This suggests that an ultra-precise determination of $n(\omega)$ requires a precise determination of $\Delta\phi_{exp}(\omega)$, where $\sigma_{\Delta\phi(\omega)}$ is small enough to distinguish the small phase change associated with a change in $M$ by $\pm 1$.

In summary, we have demonstrated a method that is able to simultaneously determine $d$ and $n(\omega)$ of an undoped silicon wafer with $\sigma_d = 1.5$ nm and $\sigma_n = 1.2 \times 10^{-5}$, respectively. We determined the absolute phase spectrum by DCS and a simple criterion with respect to data smoothness. DCS can be used to measure both amplitude and phase with high frequency resolution and accuracy. The dramatic improvement in the precision of the absolute sample-induced phase change $\Delta\Phi_{abs}(\omega)$ that is achieved by our method, allows us to measure $d$ and $n(\omega)$ simultaneously with degrees of precision that were previously unachievable.

## Methods

**Dual comb spectroscopy measurement**

In this work we use DCS as described in ref[41]. We used two OFCs based on erbium(Er)-doped-fiber-based mode-locked lasers with slightly different repetition rates ($f_r$ and $f_r$-$\Delta f_r$). The two OFCs are referred to as the signal (S) comb and the local (L) comb. In our experiment, $f_r$ was ~48 MHz and $\Delta f_r$ was set to about 69 Hz. From each laser, one of the two output beams was amplified by an Er-doped fiber amplifier (EDFA) and its spectral bandwidth was broadened by a highly nonlinear fiber (HNLF). The obtained two output beams with the broadened spectrum were used to measure the interferograms. The other two output beams were used to detect the carrier-envelope offset frequencies of the S- and the L-combs, $f_{S,ceo}$ and $f_{L,ceo}$, by $f$-$2f$ interferometers. We also detected the beat notes between a 1.54-μm continuous-wave (CW) laser and one of the comb modes for both the S-comb and the L-comb ($f_{S,beat}$ and $f_{L,beat}$). $f_{S,ceo}$ and $f_{L,ceo}$ were phase-locked to the radio frequency (RF) reference signals generated by function generators. Each carrier-envelope offset frequency was phase-locked via a feedback to the current of the corresponding pump LD. $f_{S,beat}$ and $f_{L,beat}$ were also phase-locked to the RF reference signals via feedback loops (to the electro-optic phase modulator, to the piezo-electronic transducer, and to the Peltier element in each oscillator) with different time constants. We controlled the values of $f_{S,ceo}$–$f_{L,ceo}$ and $f_{S,beat}$–$f_{L,beat}$ to obtain multiples of $\Delta f_r$ in order to satisfy the condition for coherent averaging[42]. As the CW laser is stabilized to an ultra-stable cavity, the relative line width between the S-comb and the L-comb is below 1 Hz, which is the resolution limit of the spectrum analyzer.

For the data acquisition process we first moved the sample to the measurement position. Then, we measured about 60 interferograms (recording time: $60/\Delta f_r \approx 0.87$ s). The averaged interferogram is referred to as $U_{1,w/}(t)$ and contains the sample-related information of the first dataset. As 0.87 s is shorter than the coherence time of our DCS system, it is possible to average these 60 interferograms in the time domain. Then we removed the sample from the measurement position by moving the translation stage and again measured about 60 interferograms, and the averaged interferogram is referred to as $U_{1,w/o}(t)$. We repeated the whole procedure for about 10 hours to obtain 5500 pairs of $U_{w/}(t)$ and $U_{w/o}(t)$. To identify each single dataset, we use the notation $U_{i,w/}(t)$ and $U_{i,w/o}(t)$ with $i = 1$ to 5500.

**Calculation of the transmittance and phase difference spectra**

To use the sample-path and ref.-path information (e.g. the peaks at 11.2 ns and 10.3 ns in Fig. 4a) separately, we extracted 7500 data points around the maximum intensity of each interference-signal peak and replaced the other data points with zeros. When the sample is located at the measurement position, the obtained interferograms containing only the data from the sample path and the ref. path are defined as $U_{i,w/}^{sam}(t)$ and $U_{i,w/}^{ref}(t)$, respectively. When the sample has been removed by using the translation stage, the obtained interferograms are defined as $U_{i,w/o}^{sam}(t)$ and $U_{i,w/o}^{ref}(t)$, respectively. By using these definitions, the transmittance $T_i(\omega)$ and the phase difference $\Delta\phi_{i,exp}(\omega)$ can be written as follows:

$$T_i(\omega) = \frac{\left|F\{U_{i,w/}^{sam}(t)\}\right|^2}{\left|F\{U_{i,w/o}^{sam}(t)\}\right|^2} \cdot \frac{\left|F\{U_{i,w/o}^{ref}(t)\}\right|^2}{\left|F\{U_{i,w/}^{ref}(t)\}\right|^2} \tag{4}$$

$$\Delta\phi_{i,exp}(\omega) = \left[\arg\{F(U_{i,w/}^{sam}(t))\} - \arg\{F(U_{i,w/}^{ref}(t))\}\right] - \left[\arg\{F(U_{i,w/o}^{sam}(t))\} - \arg\{F(U_{i,w/o}^{ref}(t))\}\right]$$

(5)

Here, $F$ means Fourier transform. As shown in equations (4) and (5), we use the ref.-path information to cancel fluctuations in our optical setup, and we also use it as the reference phase.

**Calculation of the complex refractive index**

For the calculation of the complex refractive index $N(\omega)$, we require $T(\omega)$, $\Delta\phi(\omega)$, $M$, and $d$. Thus, $N(\omega)$ depends on $\Delta\Phi_{abs}(\omega)$ since $\Delta\Phi_{abs}(\omega) = 2\pi M + \Delta\phi(\omega)$, where $\Delta\phi(\omega)$ is the unwrapped curve of the measured phase difference defined in equation (5).

$T(\omega)$ and $\Delta\Phi_{abs}(\omega)$ are connected with $N(\omega)=n(\omega)+ik(\omega)$ and $d$ as shown in the following equation[33]:

$$\sqrt{T(\omega)}\exp[i\Delta\Phi_{abs}(\omega)] = \frac{t_{as}(\omega)t_{sa}(\omega)}{1-r_{sa}(\omega)^2\exp\left[i\frac{2N(\omega)d}{c}\omega\right]}\exp\left[i\frac{(N(\omega)-n_{air})d}{c}\omega\right] \quad (6)$$

According to Edlén's formula[43], $n_{air} = 1.000266$ around 195 THz. $c$ is the speed of light in vacuum. $t_{as}(t_{sa})$ is the amplitude transmittance coefficient for transmittance from air to the sample (from the sample to air). $r_{sa}$ is the amplitude reflectance coefficient at the sample/air interface for light inside the sample. According to the Fresnel's law, these coefficients can be written as:

$$t_{as}(\omega) = \frac{2n_{air}}{N(\omega)+n_{air}} \quad (7)$$

$$t_{sa}(\omega) = \frac{2N(\omega)}{N(\omega)+n_{air}} \quad (8)$$

$$r_{sa}(\omega) = \frac{N(\omega)-n_{air}}{N(\omega)+n_{air}} \quad (9)$$

Using equation (6)–(9) we obtain the following expressions for the real and imaginary parts of $N(\omega)$:

$$n(\omega) = \frac{c}{\omega d}\left[2\pi M + \Delta\phi(\omega) + \frac{n_{air}d}{c}\omega - \text{Arg}(t_{as}(\omega)t_{sa}(\omega)) + \text{Arg}\left\{1-r_{sa}^2\exp\left(i\frac{2N(\omega)d}{c}\omega\right)\right\}\right] \quad (10)$$

$$k(\omega) = -\frac{c}{2\omega d}\log\left(T(\omega)\cdot\left|1-r_{sa}^2(\omega)\exp\left[i\frac{2N(\omega)d}{c}\omega\right]\right|^2\cdot|t_{as}(\omega)t_{sa}(\omega)|^{-2}\right) \quad (11)$$

Here, we added the term for multiple internal reflections, $1-r_{sa}^2\exp\left(i\frac{2Nd}{c}\omega\right)$, to the expressions of $n(\omega)$ and $k(\omega)$ derived by Tripathi et al.[44]. In our calculations, we simultaneously solve equations (10) and (11) by using the fsolve function of the software MATLAB.

**Characterization of the levels of smoothness of *n(ω)* and *k(ω)***

For a correct characterization of the smoothness, it is important that the evaluation function only considers the artificially induced oscillating component in *n(ω)* and *k(ω)*, that is, the oscillating behavior as a function of frequency that is caused by an incorrect analysis of the effect of multiple internal reflections in the *T(ω)* and *Δϕ(ω)* spectra. In undoped silicon, *n(ω)* linearly increases with *ω* in the measure frequency range. If we use incorrect values for *M* and *d*, *n(ω)* increases with *ω* but also contains a superimposed oscillating component, i.e. *n(ω)* can be expressed as $n(\omega)=\alpha+\beta\omega+f(\omega)$, where *α* and *β* are constants and *f(ω)* is the oscillating component. Therefore, by using the second-order derivative, we can evaluate *f(ω)*. Here, we consider that the second-order derivative is sufficient to evaluate the levels of smoothness of *n(ω)* and *k(ω)*, because the second-order derivatives of *n(ω)* and *k(ω)* oscillate around zero. This means that all components except the oscillating component have been successfully removed from the data.

**Authors' contributions**

K.S. and S.W. conceived the study, and K.S., S.O., and M.O. performed the experiments. K.S., M.O., and H.I. analyzed the data. K.S., M.O., and S.W. co-wrote the manuscript, and all authors provided critical comments on the manuscript.

**Acknowledgement**

This work was partially supported by JSPS KAKENHI Grant Nos. JP18H02040, JP23560048, and JP26800217, and JST CREST Grant No. JPMJCR19J4, Japan.